\documentclass[onecolumn,11pt]{article}
\usepackage[top=1in, bottom=1in, left=1in, right=1in]{geometry}
\usepackage{amsmath,amsfonts,amscd,amssymb}
\usepackage{graphicx}
\usepackage{epstopdf}
\usepackage{overpic}
\usepackage{cancel}
\usepackage{rotating}
\usepackage{url}
\usepackage{caption}
\usepackage{color}
\usepackage{rotating}
\usepackage{multirow}
\usepackage{wrapfig}
\usepackage{mathtools}
\usepackage{subeqnarray}
\usepackage{setspace}
\usepackage{palatino} 
\usepackage[numbers,sort&compress]{natbib}

\usepackage[bottom,flushmargin,hang,multiple]{footmisc}
\usepackage{lipsum}
\newcommand\blfootnote[1]{%
  \begingroup
  \renewcommand\thefootnote{}\footnote{#1}%
  \addtocounter{footnote}{-1}%
  \endgroup
}

\newcommand{\beginsupplement}{%
        \setcounter{section}{0}
        \renewcommand{\thesection}{S\arabic{section}}%
        \setcounter{table}{0}
        \renewcommand{\thetable}{S\arabic{table}}%
        \setcounter{figure}{0}
        \renewcommand{\thefigure}{S\arabic{figure}}%
     }

\definecolor{header1}{cmyk}{0,0,0,1}

\newcommand{\Wv}{\mathbf{W}}
\newcommand{\Xv}{\mathbf{X}}
\newcommand{\Zv}{\mathbf{Z}}
\newcommand{\bv}{\mathbf{b}}
\newcommand{\fv}{\mathbf{f}}
\newcommand{\gv}{\mathbf{g}}
\newcommand{\lv}{\mathbf{l}}
\newcommand{\xv}{\mathbf{x}}
\newcommand{\zv}{\mathbf{z}}
\newcommand{\Thetav}{\boldsymbol{\Theta}}
\newcommand{\Upsilonv}{\boldsymbol{\Upsilon}}
\newcommand{\Xiv}{\boldsymbol{\Xi}}
\newcommand{\thetav}{\boldsymbol{\theta}}
\newcommand{\xiv}{\boldsymbol{\xi}}

\newcommand\ddt{\frac{d}{dt}}

\newcommand\Rb{\mathbb{R}}

\newcommand{\Lrecon}{\mathcal{L}_{\text{recon}}}
\newcommand{\Lsindyx}{\mathcal{L}_{d\xv/dt}}
\newcommand{\Lsindyz}{\mathcal{L}_{d\zv/dt}}
\newcommand{\Lreg}{\mathcal{L}_{\text{reg}}}

\title{\LARGE{\vspace{-.55in}\textbf{Simultaneous discovery of coordinates and parsimonious dynamics: SINDy autoencoders}}\vspace{-.175in}}
\title{\vspace{-.55in}{\fontsize{16}{16}\selectfont \textbf{Data-driven discovery of coordinates and governing equations}}\vspace{-.15in}}

\author{\normalsize{Kathleen Champion$^{1*}$, Bethany Lusch$^2$, J. Nathan Kutz$^1$, Steven L. Brunton$^3$}\\
\footnotesize{$^1$ Department of Applied Mathematics, University of Washington, Seattle, WA 98195, United States}\\
\footnotesize{$^2$ Argonne National Laboratory, Lemont, IL 60439, United States}\\
\footnotesize{$^3$ Department of Mechanical Engineering, University of Washington, Seattle, WA 98195, United States\vspace{-.2in}}
}
\date{}
\begin{document}
\maketitle

\blfootnote{$^*$ Corresponding author (kpchamp@uw.edu).}
\vspace{-.2in}
\begin{abstract}
The discovery of governing equations from scientific data has the potential to transform data-rich fields that lack well-characterized quantitative descriptions. 
Advances in sparse regression are currently enabling the tractable identification of both the structure and parameters of a nonlinear dynamical system from data. 
The resulting models have the fewest terms necessary to describe the dynamics, balancing model complexity with descriptive ability, and thus promoting interpretability and generalizability.  This provides an algorithmic approach to Occam's razor for model discovery.
However, this approach fundamentally relies on an effective coordinate system in which the dynamics have a simple representation. 
In this work, we design a custom autoencoder to discover a coordinate transformation into a reduced space where the dynamics may be sparsely represented.  
Thus, we simultaneously learn the governing equations and the associated coordinate system.  
We demonstrate this approach on several example high-dimensional dynamical systems with low-dimensional behavior.  
The resulting modeling framework combines the strengths of deep neural networks for flexible representation and sparse identification of nonlinear dynamics (SINDy) for parsimonious models.  It is the first method of its kind to place the discovery of coordinates and models on an equal footing. \\ 

\end{abstract}

\section{Introduction}

Governing equations are of fundamental importance across all scientific disciplines.   Accurate models allow for fundamental understanding of physical processes, which in turn gives rise to an infrastructure for the development of technology.   The traditional derivation of governing equations is based on underlying first principles, such as conservation laws and symmetries, or from universal laws, such as gravitation.  However, in many modern systems, governing equations are unknown or only partially known, and recourse to first principles derivations is untenable.  On the other hand, many of these modern systems have rich time-series data due to the emergence of sensor and measurement technologies (e.g. in biology and climate science).  This has given rise to the new paradigm of data-driven model discovery.  Indeed, the data-driven discovery of dynamical systems is the focus of intense research efforts across the physical and engineering sciences~\cite{bongard_automated_2007,Yao2007physicad,schmidt_distilling_2009,Rowley2009jfm,schmid_dynamic_2010,Wang2011prl,Benner2015siamreview,peherstorfer2016data,brunton_discovering_2016,Kutz2016book,Rudy2017sciadv,Yair2017pnas,Duraisamy2018arfm,pathak2018model,battaglia2018relational}.  
A central tension in model discovery is the balance between model efficiency and descriptive capabilities. 
Parsimonious models strike this balance, having the fewest terms required to capture essential interactions, and not more~\cite{bongard_automated_2007,schmidt_distilling_2009,brunton_discovering_2016,Rudy2017sciadv}.  
Related to Occam's razor, parsimonious models tend to be more interpretable and generalizable. 
However, obtaining parsimonious models is fundamentally linked to the coordinate system in which the dynamics are measured. 
Without knowledge of the proper coordinates, standard approaches may fail to discover simple dynamical models.  
In this work, we develop a machine learning approach that simultaneously discovers effective coordinates via a custom autoencoder~\cite{baldi1989neural,goodfellow2016deep,lusch2018deep}, along with the parsimonious dynamical system model via sparse regression in a library of candidate terms~\cite{brunton_discovering_2016}.    The joint discovery of models and coordinates is critical for understanding many modern complex systems.

%
%
Numerous recent approaches leverage machine learning to identify dynamical systems models from data~\cite{Crutchfield1987cs,kevrekidis_equation-free_2003,Schmidt2011pb,Daniels2015naturecomm,Daniels2015plosone,Sugihara2012science,Roberts2014book,Ye2015pnas}, with many using neural networks to model time-series data~\cite{gonzalez1998identification,Mardt2017arxiv,vlachas_data-driven_2018,wehmeyer2018time,yeung_learning_2017,Takeishi2017nips,lusch2018deep,otto2017linearly,raissi2017hidden,raissi2017physics1,raissi2017physics2,raissi2018multistep,bar2018data}. 
When interpretability and generalizability are primary concerns, it is important to identify \emph{parsimonious} models that have the fewest terms required to describe the dynamics, which is the antithesis of neural networks whose parametrizations are exceedingly large.
A breakthrough approach used symbolic regression to learn the form of dynamical systems and governing laws from data~\cite{bongard_automated_2007,schmidt_distilling_2009,cornforth_symbolic_2012}. 
Sparse identification of nonlinear dynamics (SINDy)~\cite{brunton_discovering_2016} is a related approach that uses sparse regression to find the fewest terms in a library of candidate terms required to model a dynamical system.  
Because this approach is based on a sparsity-promoting linear regression, it is possible to incorporate partial knowledge of the physics, such as symmetries, constraints, and conservation laws~\cite{Loiseau2017jfm}. 
%
%
Successful model identification relies on the assumption that the dynamics are measured in a coordinate system in which the dynamics may be sparsely represented.  
While simple models may exist in one coordinate system, a different coordinate system may obscure these parsimonious representations.  
For modern applications of data-driven discovery, there is no reason to believe that our sensors are measuring the correct variables to admit a parsimonious representation of the dynamics.
This motivates the systematic and automated discovery of coordinate transformations to facilitate this sparse representation, which is the subject of this work.

The challenge of discovering an effective coordinate system is as fundamental and important as model discovery.  
Many key historical scientific breakthroughs were enabled by the discovery of appropriate coordinate systems.  
Celestial mechanics, for instance, was revolutionized by the heliocentric coordinate system of Copernicus, Galileo, and Kepler, thus displacing Ptolemy's {\em doctrine of the perfect circle}, which was dogma for more than a millennium.  
Fourier introduced his famous transform to simplify the representation of the heat equation, resulting in a \emph{sparse}, diagonal, decoupled linear system.  
Eigen-coordinates have been used more broadly to enable simple and {sparse} decompositions, for example in quantum mechanics and electrodynamics, to characterize energy levels in atoms and propagating modes in waveguides, respectively.  
Principal component analysis (PCA) is one of the most prolific modern coordinate discovery methods, representing high-dimensional data in a low-dimensional linear subspace~\cite{Pearson:1901,HLBR_turb}.  
%
Nonlinear extensions of PCA have been enabled by a neural network architecture, called an autoencoder~\cite{baldi1989neural,Milano2002jcp,goodfellow2016deep}. 
However, PCA coordinates and autoencoders generally do not take dynamics into account and, thus, may not provide the right basis for parsimonious dynamical models. 
In a related vein, Koopman analysis seeks to discover coordinates that linearize nonlinear dynamics~\cite{koopman_hamiltonian_1931,mezic_spectral_2005,budisic_applied_2012,mezic_analysis_2013}; while linear models are useful for prediction and control, they cannot capture the full behavior of many nonlinear systems. 
Thus, it is important to develop methods that combine simplifying coordinate transformations and nonlinear dynamical models.
We advocate for a balance between these approaches, identifying coordinate transformations where only a few nonlinear terms are present, as in the classic theory of near-identity transformations and normal forms~\cite{guckenheimer_holmes,Yair2017pnas}. 

In this work we present a method for discovery of nonlinear coordinate transformations that enable associated parsimonious dynamics. Our method combines a custom autoencoder network with a SINDy model for parsimonious nonlinear dynamics. The autoencoder architecture enables the discovery of reduced coordinates from high-dimensional input data that can be used to reconstruct the full system. The reduced coordinates are found along with nonlinear governing equations for the dynamics in a joint optimization. We demonstrate the ability of our method to discover parsimonious dynamics on three examples: a high-dimensional spatial data set with dynamics governed by the chaotic Lorenz system, a spiral wave resulting from the reaction-diffusion equation, and the nonlinear pendulum.
These results demonstrate how to focus neural networks to discover interpretable dynamical models.  
Critically, the proposed method is the first to provide a mathematical framework that places the discovery of coordinates and models on equal footing.

\section{Background}\label{sec:background}

\subsection{Sparse identification of nonlinear dynamics}

We review the {\em sparse identification of nonlinear dynamics} (SINDy) \cite{brunton_discovering_2016} algorithm, which is a mathematical regression technique for extracting parsimonious dynamics from time-series data.   The method takes snapshot data $\xv(t) \in \Rb^n$ and attempts to discover a best-fit dynamical system with as few terms as possible:
%
\begin{equation}
  \ddt \xv(t) = \fv(\xv(t)).
    \label{eq:dynamical_system_x}
\end{equation}
The snapshots represent measurements of the state of the system in time $t$, and the function $\fv$ constrains how the dynamics of the system evolve in time. We seek a parsimonious model for the dynamics, resulting in a function $\fv$ that contains only a few active terms: it is sparse in a basis of possible functions.  This is consistent with our extensive knowledge of a diverse set of evolution equations used throughout the physical, engineering and biological sciences.   Thus, the functions that comprise $\fv$ are typically known from our extensive modeling experience.

SINDy frames model discovery as a sparse regression problem. Assuming snapshot derivatives are available, or can be calculated from data, the snapshots are stacked to form data matrices
\begin{equation}
  \Xv = \left(\begin{array}{cccc}
    x_1(t_1) & x_2(t_1) & \cdots & x_n(t_1) \\
    x_1(t_2) & x_2(t_2) & \cdots & x_n(t_2) \\
    \vdots & \vdots & \ddots & \vdots \\
    x_1(t_m) & x_2(t_m) & \cdots & x_n(t_m) \\
  \end{array}\right), \quad
  \dot{\Xv} = \left(\begin{array}{cccc}
    \dot{x}_1(t_1) & \dot{x}_2(t_1) & \cdots & \dot{x}_n(t_1) \\
    \dot{x}_1(t_2) & \dot{x}_2(t_2) & \cdots & \dot{x}_n(t_2) \\
    \vdots & \vdots & \ddots & \vdots \\
    \dot{x}_1(t_m) & \dot{x}_2(t_m) & \cdots & \dot{x}_n(t_m) \\
  \end{array}\right). \nonumber
\end{equation}
with $\Xv,\dot{\Xv} \in \Rb^{m \times n}$. Although $\fv$ is unknown, we can construct an extensive library of $p$ potential candidate basis functions $\Thetav(\Xv) = [\thetav_1(\Xv) \cdots \thetav_p(\Xv)] \in \Rb^{m\times p}$, where each $\thetav_j$ is a candidate basis function or model term.  
We assume $m\gg p$ so that the number of data snapshots is much larger than the number of candidate library functions; it may be necessary to sample transient dynamics and multiple initial conditions to improve the condition number of $\Thetav$. 
The choice of basis functions typically reflects some knowledge about the system of interest: a common choice is polynomials in $\xv$ as these are elements of many canonical models of dynamical systems~\cite{guckenheimer_holmes}. The library is used to formulate a regression problem to approximately solve the overdetermined linear system of equations
\begin{equation}
  \dot{\Xv} = \Thetav(\Xv)\Xiv \nonumber
\end{equation}
where the unknown matrix $\Xiv = (\xiv_1\ \xiv_2\ \cdots\ \xiv_n)\in \Rb^{p\times n}$ is the set of coefficients that determine the active terms from $\Thetav(\Xv)$ in the dynamics $\fv$. Sparsity-promoting regression is used to find parsimonious models, ensuring that $\Xiv$,  or more precisely each $\xiv_j$, is sparse and only a few columns of $\Thetav(\Xv)$ are selected.  For high-dimensional systems, the goal is to identify a low-dimensional state $\zv=\varphi(\xv)$ with dynamics $\dot{\zv} = \gv(\zv)$, as in Sec.~\ref{sec:methods}. The standard SINDy approach uses a sequentially thresholded least squares algorithm to find the coefficients \cite{brunton_discovering_2016}, which is a proxy for $\ell_0$ optimization~\cite{zheng2019unified} and has convergence guarantees~\cite{Zhang2018arxiv}.  
Yao and Bollt~\cite{Yao2007physicad} previously formulated the dynamical system identification problem as a similar linear inverse problem, although sparsity-promoting regression was not used, so the resulting models generally included all terms in $\boldsymbol{\Theta}$.   In either case, an appealing aspect of this model discovery formulation is that it results in an overdetermined linear system of equations for which many regularized solution techniques exist.  Thus, it provides a computationally efficient counterpart to other model discovery frameworks~\cite{schmidt_distilling_2009,Wang2011prl}. 

SINDy has been widely applied to identify models for fluid flows~\cite{Loiseau2017jfm,Loiseau2018jfm}, optical systems~\cite{Sorokina2016oe}, chemical reaction dynamics~\cite{Hoffmann2018arxiv},  convection in a plasma~\cite{Dam2017pf}, structural modeling~\cite{lai2019sparse}, and for model predictive control~\cite{Kaiser2018prsa}. 
There are also a number of theoretical extensions to the SINDy framework, including for identifying partial differential equations~\cite{Rudy2017sciadv,Schaeffer2017prsa},  multiscale physics~\cite{champion2019discovery}, parametrically dependent dynamical models~\cite{rudy2018data},  hybrid (switching) dynamical systems~\cite{mangan2019model}, and models with rational function nonlinearities~\cite{Mangan2016ieee}.  It can also incorporate partially known physics and constraints~\cite{Loiseau2017jfm} and identify models based on physically realistic sensor measurements~\cite{Loiseau2018jfm}.  The algorithm can also be reformulated to include integral terms for noisy data~\cite{Schaeffer2017pre} or handle incomplete or limited data~\cite{Tran2016arxiv,schaeffer2018extracting}.   The selected modes can also be evaluated using information criteria for model selection~\cite{Mangan2017prsa}.    These diverse mathematical developments provide a mature framework for broadening the applicability of the model discovery method.

\subsection{Neural networks for dynamical systems}
The success of neural networks (NNs) on problems such as image classification \cite{krizhevsky_imagenet_2012} and speech recognition~\cite{hinton2012deep} has led to the use of NNs to perform a wide range of tasks in science and engineering. One recent area of focus has been the use of NNs for studying dynamical systems, which has a surprisingly rich history~\cite{gonzalez1998identification}.  In addition to improving solution techniques for systems with known equations \cite{raissi2017hidden,raissi2017physics1,raissi2017physics2,raissi2018multistep,bar2018data}, deep learning has been used for understanding and predicting dynamics for complex systems with potentially unknown equations \cite{Mardt2017arxiv,vlachas_data-driven_2018,wehmeyer2018time,yeung_learning_2017,Takeishi2017nips,lusch2018deep,otto2017linearly}.
%
Several recent methods have trained NNs to predict dynamics, including a time-lagged autoencoder which takes the state at time $t$ as input data and uses an autoencoder-like structure to predict the state at time $t + \tau$ \cite{wehmeyer2018time}. Other approaches use a recurrent NN architecture,  particularly long short-term memory (LSTM) networks, for applications involving sequential data \cite{hochreiter1997long,hochreiter1997lstm,lipton2015critical}. 
 LSTMs have recently been used to perform forecasting on chaotic dynamical systems \cite{vlachas_data-driven_2018}.  Reservoir computing has also enabled impressive predictions~\cite{pathak2018model}. 
Autoencoders are increasingly being leveraged for dynamical systems because of their close relationship to other dimensionality reduction techniques~\cite{Milano2002jcp,carlberg2018recovering,gonzalez2018learning,lee2018model}.

Another class of NNs use deep learning to discover coordinates for Koopman analysis. Koopman theory seeks to discover coordinates that linearize nonlinear dynamics \cite{koopman_hamiltonian_1931,mezic_spectral_2005,budisic_applied_2012,mezic_analysis_2013}. Methods such as {\em dynamic mode decomposition} (DMD)~\cite{schmid_dynamic_2010,Rowley2009jfm,schmid2011applications,Kutz2016book}, extended DMD~\cite{williams_datadriven_2015}, kernel DMD~\cite{williams_kernel-based_2015}, and time-delay DMD~\cite{brunton_chaos_2017,Arbabi2016arxiv} build linear models for dynamics, but these methods rely on a proper set of coordinates for linearization.  
Several recent works have focused on the use of deep learning methods to discover the proper coordinates for DMD and extended DMD \cite{li_extended_2017,yeung_learning_2017,Takeishi2017nips}. Other methods seek to learn Koopman eigenfunctions and the associated linear dynamics directly using autoencoders~\cite{otto2017linearly,lusch2018deep}.

Despite their widespread use, NNs face three major challenges:  generalization, extrapolation, and interpretation.   The hallmark success stories of NNs (computer vision and speech, for instance) have been on data sets that are fundamentally interpolatory in nature.  The ability to extrapolate, and as a consequence generalize, is known to be an underlying weakness of NNs.  This is especially relevant for dynamical systems and forecasting, which is typically an extrapolatory problem by nature.  Thus models trained on historical data will generally fail to predict future events that are not represented in the training set.  An additional limitation of deep learning is the lack of interpretability of the resulting models. While attempts have been made to interpret NN weights, network architectures are typically complicated with the number of parameters (or weights) far exceeding the original dimension of the dynamical system. The lack of interpretability also makes it difficult to generalize models to new data sets and parameter regimes. However, NN methods still have the potential to learn general, interpretable dynamical models if properly constrained or regularized. In addition to methods for discovering linear embeddings \cite{lusch2018deep,otto2017linearly}, deep learning has also been used for parameter estimation of PDEs \cite{raissi2017physics2,raissi2018multistep}.

\section{SINDy Autoencoders}\label{sec:methods}

\begin{figure*}
\vspace{-.1in}
 \centering
\begin{overpic}[width=.95\textwidth]{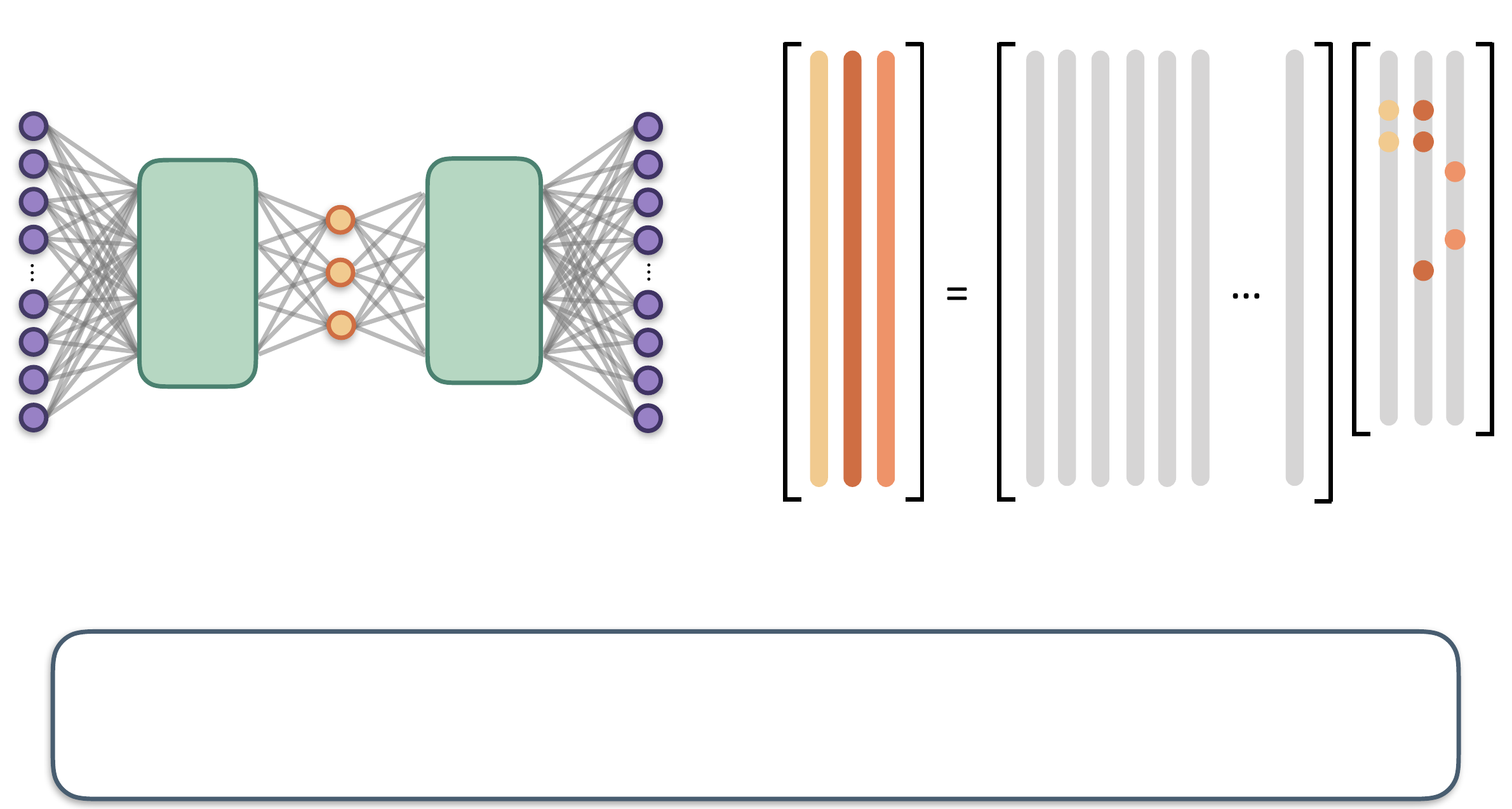}
  \put(1,51){(a)}
  \put(48,51){(b)}
  \put(1,21){\large $\xv(t)$}
  \put(21.25,21){\large $\zv(t)$}
  \put(40.5,21){\large $\hat{\xv}(t)$}
  \put(12,35){\large $\varphi$}
  \put(31,35){\large $\psi$}
  \put(54,51.5){\small $\dot{z}_1$}
  \put(56,51.5){\small $\dot{z}_2$}
  \put(58,51.5){\small $\dot{z}_3$}
  \put(68.2,51.5){\small $1$}
  \put(69.8,51.5){\small $z_1$}
  \put(72,51.5){\small $z_2$}
  \put(74.2,51.5){\small $z_3$}
  \put(76.4,51.5){\small $z_1^2$}
  \put(78.6,51.5){\small $z_1 z_2$}
  \put(85,51.5){\small $z_3^3$}
  \put(91.2,51.5){\small $\xi_1$}
  \put(93.3,51.5){\small $\xi_2$}
  \put(95.4,51.5){\small $\xi_3$}
  \put(55,17.5){\large $\dot{\Zv}$}
  \put(74,17.5){\large $\Thetav(\Zv)$}
  \put(93,17.5){\large $\Xiv$}
  \put(48,14){$\dot{\zv}_i = \nabla_\xv\varphi(\xv_i)\dot{\xv}_i$}
  \put(68,14){$\Thetav(\zv_i^T) = \Thetav(\varphi(\xv_i)^T)$}
  \put(9,8){\small$\underbrace{\left\| \xv - \psi(\zv)\right\|_2^2} + \underbrace{\lambda_1 \left\|\dot{\xv} - \left(\nabla_\zv\psi(\zv)\right)\left(\Thetav(\zv^T)\Xiv\right)\right\|_2^2} + \underbrace{\lambda_2 \left\|\left(\nabla_\xv\zv\right)\dot{\xv} - \Thetav(\zv^T)\Xiv\right\|_2^2} + \underbrace{\lambda_3 \left\| \Xiv \right\|_1}$}
  \put(5.5,3.2){\small reconstruction loss}
  \put(31,3.2){\small SINDy loss in $\dot{\xv}$}
  \put(59,3.2){\small SINDy loss in $\dot{\zv}$}
  \put(79.5,2.7){\small \parbox{6em}{\centering SINDy\\regularization}}
\end{overpic}
\caption{Schematic of the SINDy autoencoder method for simultaneous discovery of coordinates and parsimonious dynamics. (a) An autoencoder architecture is used to discover intrinsic coordinates $\zv$ from high-dimensional input data $\xv$. The network consists of two components: an encoder $\varphi(\xv)$, which maps the input data to the intrinsic coordinates $\zv$, and a decoder $\psi(\zv)$, which reconstructs $\xv$ from the intrinsic coordinates. (b) A SINDy model captures the dynamics of the intrinsic coordinates. The active terms in the dynamics are identified by the nonzero elements in $\Xiv$, which are learned as part of the NN training. The time derivatives of $\zv$ are calculated using the derivatives of $\xv$ and the gradient of the encoder $\varphi$. The inset shows the pointwise loss function used to train the network. The loss function encourages the network to minimize both the autoencoder reconstruction error and the SINDy loss in $\zv$ and $\xv$. $L_1$ regularization on $\Xiv$ is also included to encourage parsimonious dynamics.}
\label{fig:overview}
\end{figure*}
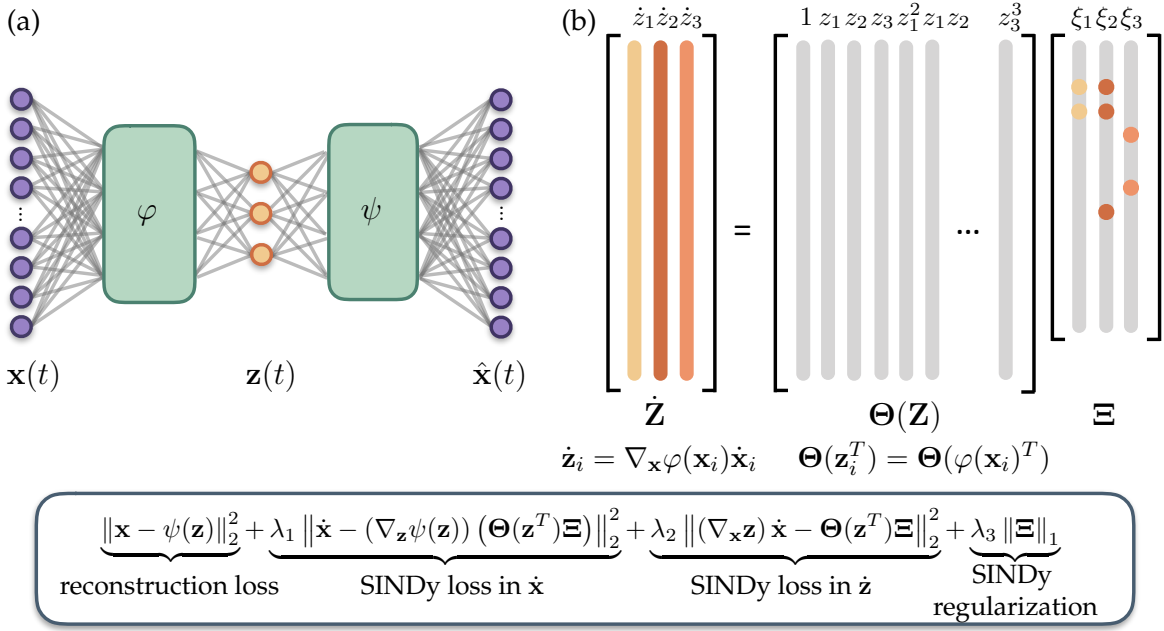

We present a method for the simultaneous discovery of sparse dynamical models and coordinates that enable these simple representations.  Our aim is to leverage the parsimony and interpretability of SINDy with the universal approximation capabilities of deep neural networks~\cite{hornik1989multilayer} in order to produce interpretable and generalizable models capable of extrapolation and forecasting.  Our approach combines a SINDy model and a deep autoencoder network to perform a joint optimization that discovers intrinsic coordinates which have an associated parsimonious nonlinear dynamical model. The architecture is shown in Figure~\ref{fig:overview}. We again consider dynamical systems of the form (\ref{eq:dynamical_system_x}).
While this dynamical model may be dense in terms of functions of the original measurement coordinates $\xv$, our method seeks a set of reduced coordinates $\zv(t)=\varphi(\xv(t)) \in \Rb^d$ ($d\ll n$) with an associated dynamical model
\begin{equation}
    \ddt\zv(t) = \gv(\zv(t))
    \label{eq:dynamical_system_z}
\end{equation}
that provides a parsimonious description of the dynamics. This means that $\gv$ contains only a few active terms. Along with the dynamical model, the method provides coordinate transforms $\varphi,\psi$ that map the measurement coordinates to intrinsic coordinates via $\zv = \varphi(\xv)$ (encoder) and back via $\xv \approx \psi(\zv)$ (decoder).

The coordinate transformation is achieved using an autoencoder network architecture. The autoencoder is a feedforward neural network with a hidden layer that represents the intrinsic coordinates. Rather than performing a task such as prediction or classification, the network is trained to output an approximate reconstruction of its input, and the restrictions placed on the network architecture (e.g. the type, number, and size of the hidden layers) determine the properties of the intrinsic coordinates \cite{goodfellow2016deep}; these networks are known to produce nonlinear generalizations of PCA~\cite{baldi1989neural}. A common choice is that the dimensionality of the intrinsic coordinates $\zv$, determined by the number of units in the corresponding hidden layer, is much lower than that of the input data $\xv$: in this case, the autoencoder learns a {\em nonlinear} embedding into a reduced latent space. Our network takes measurement data $\xv(t) \in \Rb^{n}$ from a dynamical system as input and learns intrinsic coordinates $\zv(t) \in \Rb^{d}$, where $d\ll n$ is chosen as a hyperparameter prior to training the network. 

While autoencoders can be trained in isolation to discover useful coordinate transformations and dimensionality reductions, there is no guarantee that the intrinsic coordinates learned will have associated sparse dynamical models. We require the network to learn coordinates associated with parsimonious dynamics by simultaneously learning a SINDy model for the dynamics of the intrinsic coordinates $\zv$. This regularization is achieved by constructing a library $\Thetav(\zv) = [\thetav_1(\zv), \thetav_2(\zv), \dots, \thetav_p(\zv)]$ of candidate basis functions, e.g. polynomials, and learning a sparse set of coefficients $\Xiv = [\xiv_1, \dots, \xiv_d]$ that defines the dynamical system
\begin{equation}
    \ddt\zv(t) = \gv(\zv(t)) = \Thetav(\zv(t))\Xiv.
\end{equation}
While the functions in the library must be specified prior to training, the coefficients $\Xiv$ are learned along with the NN parameters as part of the training procedure. Assuming derivatives $\dot{\xv}(t)$ of the original states are available or can be computed, one can calculate the derivative of the encoder variables as $\dot{\zv}(t) = \nabla_\xv\varphi(\xv(t))\dot{\xv}(t)$ and enforce accurate prediction of the dynamics by incorporating the following term into the loss function:
\begin{equation}
    \label{eq:loss_sindy_z}
    \Lsindyz = \left\|\nabla_\xv\varphi(\xv)\dot{\xv} - \Thetav(\varphi(\xv)^T)\Xiv\right\|_2^2.
\end{equation}
This term uses the SINDy model along with the gradient of the encoder to encourage the learned dynamical model to accurately predict the time derivatives of the encoder variables. We include an additional term in the loss function that ensures SINDy predictions can be used to reconstruct the time derivatives of the original data:
\begin{equation}
    \label{eq:loss_sindy_x}
    \Lsindyx  = \left\|\dot{\xv} - \left(\nabla_\zv\psi(\varphi(\xv))\right)\left(\Thetav(\varphi(\xv)^T)\Xiv\right)\right\|_2^2.
\end{equation}
We combine \eqref{eq:loss_sindy_z} and \eqref{eq:loss_sindy_x} with the standard autoencoder loss function
\begin{equation}
  \Lrecon = \left\| \xv - \psi(\varphi(\xv))\right\|_2^2,
\end{equation}
which ensures that the autoencoder can accurately reconstruct the input data. We also include an $L_1$ regularization on the SINDy coefficients $\Xiv$, which promotes sparsity of the coefficients and therefore encourages a parsimonious model for the dynamics. The combination of the above four terms gives the following overall loss function:
\begin{equation}
  \Lrecon + \lambda_1 \Lsindyx + \lambda_2 \Lsindyz + \lambda_3 \Lreg ,
\end{equation}
where the scalars $\lambda_1,\lambda_2,\lambda_3$ are hyperparameters that determine the relative weighting of the three terms in the loss function.

In addition to the $L_1$ regularization, to obtain a model with only a few active terms we also incorporate sequential thresholding into the training procedure as a proxy for $L_0$ sparsity~\cite{zheng2019unified}. This technique is inspired by the original algorithm used for SINDy \cite{brunton_discovering_2016}, which combined least squares fitting with sequential thresholding to obtain a sparse dynamical model. To apply sequential thresholding during training, we specify a threshold that determines the minimum magnitude for coefficients in the SINDy model. At fixed intervals throughout the training, all coefficients below the threshold are set to zero and training resumes using only the terms left in the model. 
We train the network using the Adam optimizer \cite{DBLP:journals/corr/KingmaB14}. In addition to the loss function weightings and SINDy coefficient threshold, training requires the choice of several other hyperparameters including learning rate, number of intrinsic coordinates $d$, network size, and activation functions. Full details of the training procedure are discussed in Section~\ref{sec:si_training}.

\section{Results}

\begin{figure*}
\centering
\includegraphics[width=\linewidth]{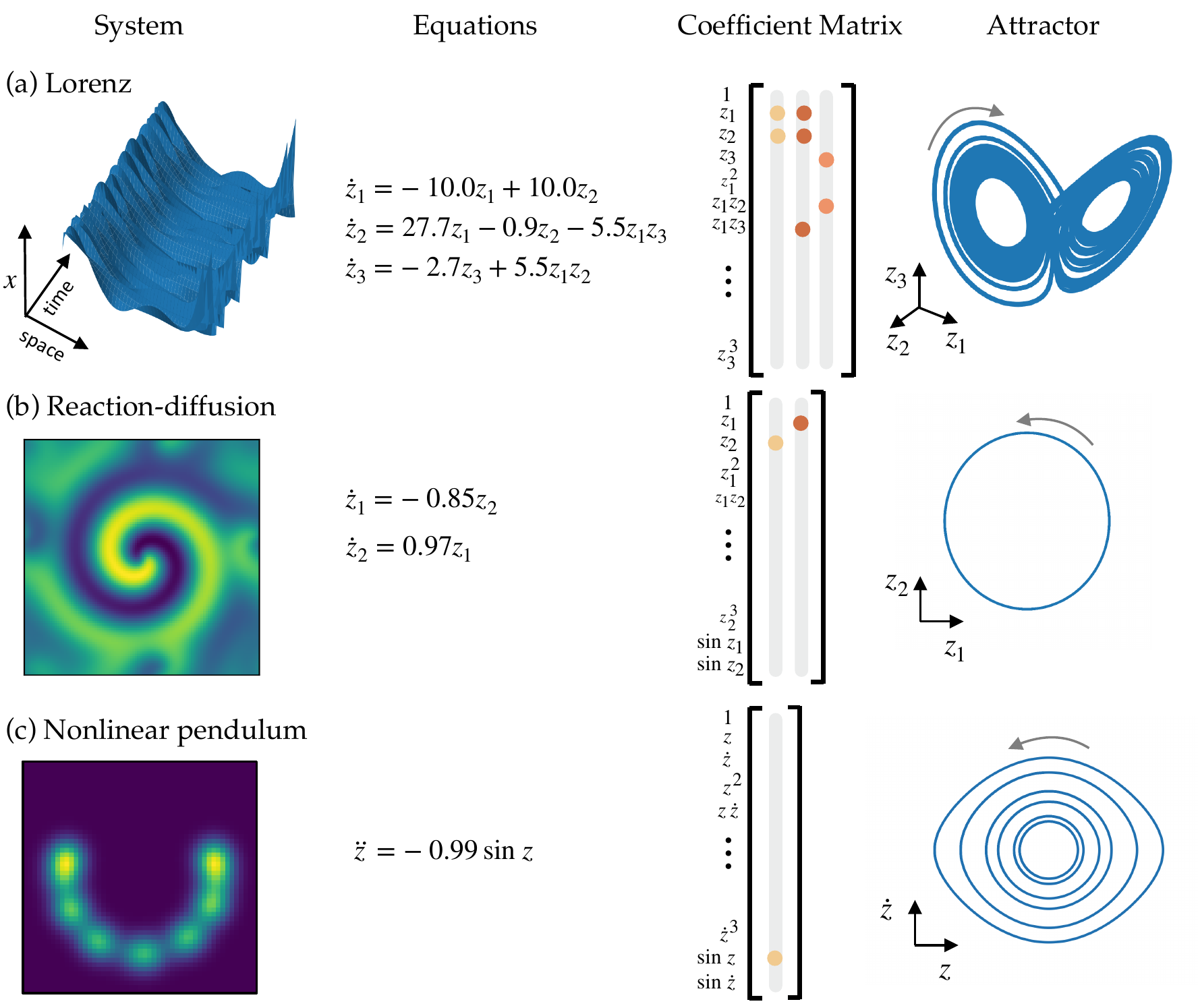}
\caption{Discovered dynamical models for example systems. (a,b,c) Equations, SINDy coefficients $\Xiv$, and attractors for the Lorenz system, reaction-diffusion system, and nonlinear pendulum.}
\label{fig:results}
\end{figure*}

We demonstrate the success of the proposed method on three example systems: a high-dimensional system with the underlying dynamics generated from the canonical chaotic Lorenz system, a two-dimensional reaction-diffusion system, and a two-dimensional  spatial representation (synthetic video) of the nonlinear pendulum.
Results are shown in Figure~\ref{fig:results}, and additional details for each example are provided in Section~\ref{sec:si_results}.

\begin{figure*}
\centering
\includegraphics[width=\linewidth]{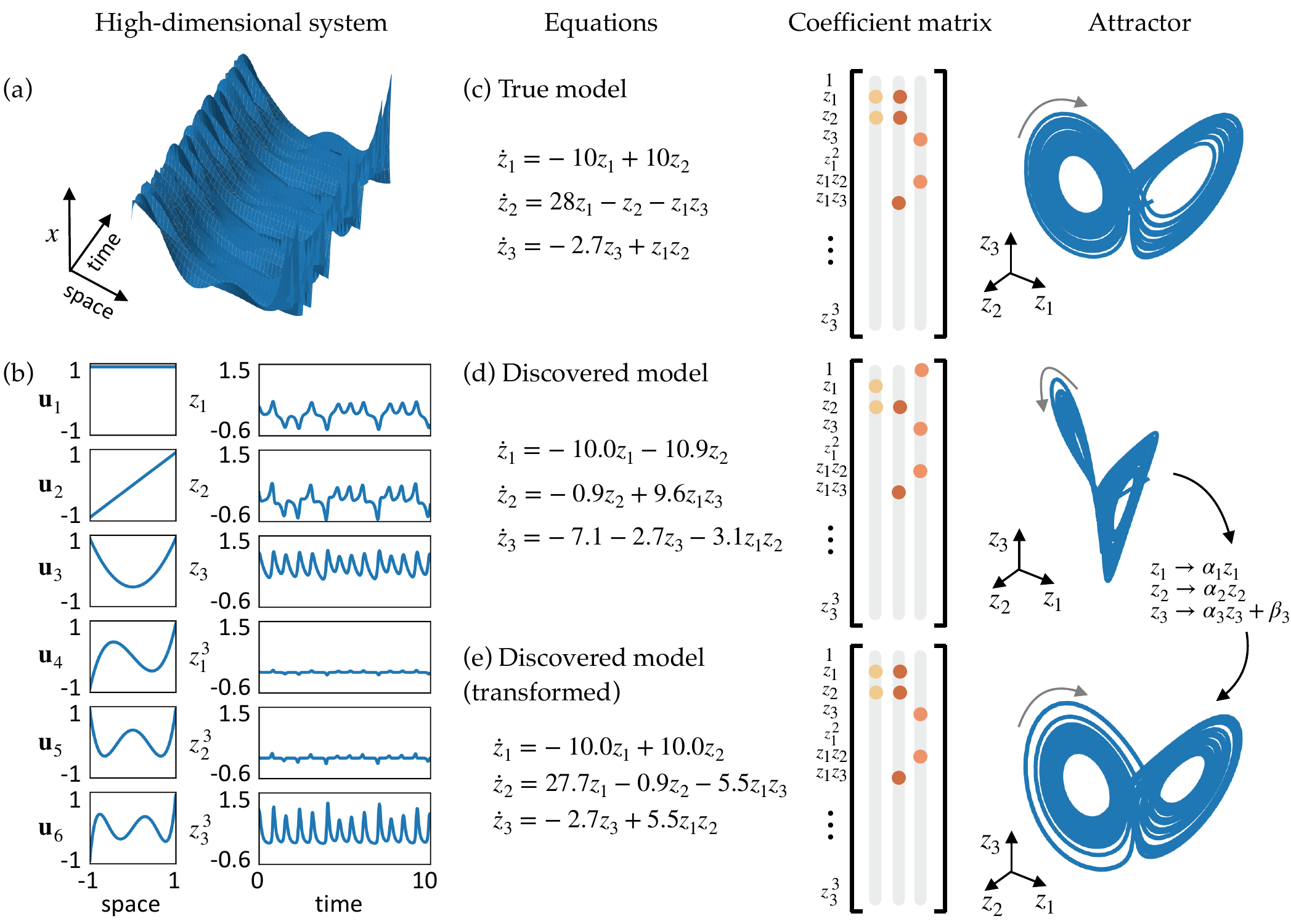}
\caption{Model results on the high-dimensional Lorenz example. (a) Trajectories of the chaotic Lorenz system ($\zv(t) \in \Rb^3$) are used to create a high-dimensional data set ($\xv(t) \in \Rb^{128}$). (b) The spatial modes are created from the first six Legendre polynomials and the temporal modes are the variables in the Lorenz system and their cubes. The spatial and temporal modes are combined to create the high-dimensional data set via \eqref{eq:lorenz_dataset}. (c,d) The equations, SINDy coefficients $\Xiv$, and attractors for the original Lorenz system and a dynamical system discovered by the SINDy autoencoder. The attractors are constructed by simulating the dynamical system forward in time from a single initial condition. (e) Applying a suitable variable transformation to the system in (d) reveals a model with the same sparsity pattern as the original Lorenz system. The parameters are close in value to the original system, with the exception of an arbitrary scaling, and the attractor has a similar structure to the original system.}
\label{fig:lorenz_detail}
\end{figure*}

\subsection{Chaotic Lorenz system}
We first construct a high-dimensional example problem with dynamics based on the chaotic Lorenz system. The Lorenz system is a canonical model used as a test case for many dynamical systems methods, with dynamics given by the following equations
\begin{subequations}
\begin{align}
  \dot{z}_1 &= \sigma(z_2 - z_1) \\
  \dot{z}_2 &= z_1(\rho - z_3) - z_2 \\
  \dot{z}_3 &= z_1 z_2 - \beta z_3.
\end{align}
\label{eq:lorenz}
\end{subequations}
The dynamics of the Lorenz system are chaotic and highly nonlinear, making it an ideal test problem for model discovery. To create a high-dimensional data set based on this system, we choose six fixed spatial modes $\mathbf{u}_1,\dots,\mathbf{u}_6 \in \Rb^{128}$, given by Legendre polynomials, and define
\begin{align}
  \mathbf{x}(t) = \mathbf{u}_1 z_1(t) + \mathbf{u}_2 z_2(t) + \mathbf{u}_3 z_3(t) + \mathbf{u}_4 z_1(t)^3 + \mathbf{u}_5 z_2(t)^3 + \mathbf{u}_6 z_3(t)^3.
  \label{eq:lorenz_dataset}
\end{align}
This results in a data set that is a nonlinear combination of the true Lorenz variables, shown in Figure~\ref{fig:lorenz_detail}a. The spatial and temporal modes that combine to give the full dynamics are shown in Figure~\ref{fig:lorenz_detail}b. Full details of how the data set is generated are given in Section~\ref{sec:si_lorenz}.

Figure~\ref{fig:lorenz_detail}d shows the dynamical system discovered by the SINDy autoencoder. While the resulting model does not appear to match the original Lorenz dynamics, the discovered model is parsimonious, with only 7 active terms, and with dynamics that live on an attractor which has a two lobe structure similar to that of the original Lorenz attractor. Additionally, by choosing a suitable variable transformation the discovered model can be rewritten in the same form as the original Lorenz system. This demonstrates that the SINDy autoencoder is able to recover the correct sparsity pattern of the dynamics. The coefficients of the discovered model are close to the original parameters of the Lorenz system, up to an arbitrary scaling, which accounts for the difference in magnitude of the coefficients of $z_1z_3$ in the second equation and $z_1z_2$ in the third equation.

On a test data set of trajectories from 100 randomly chosen initial conditions, the mean square error of the decoder reconstruction is less than $3\times 10^{-5}$ of the fraction of the variance of the input data. The fraction of the unexplained variance in predicting the derivatives $\dot{\xv}$ and $\dot{\zv}$ are $2 \times 10^{-4}$ and $7 \times 10^{-4}$, respectively. Simulations of the resulting SINDy model are able to accurately reconstruct the dynamics of a single trajectory with less than 1\% error over the duration of trajectories in the training data. Over longer durations, the trajectories start to diverge from the true trajectories. This result is not surprising due to the chaotic nature of the Lorenz system and its sensitivity to initial conditions. However, the dynamics of the discovered system match the sparsity pattern of the Lorenz system and the form of the attractor. Improved prediction over a longer duration could be achieved by increased parameter refinement or training the system with longer trajectories.

\subsection{Reaction-diffusion}

In practice, many high-dimensional data sets of interest come from dynamics governed by partial differential equations (PDEs) with more complicated interactions between spatial and temporal dynamics. To test the method on data generated by a PDE, we consider a lambda-omega reaction-diffusion system governed by
\begin{subequations}
\begin{align}
  u_t &= (1-(u^2+v^2))u + \beta (u^2+v^2) v + d_1 (u_{xx} + u_{yy}) \\
  v_t &= -\beta (u^2+v^2)u + (1-(u^2+v^2))v + d_2 (v_{xx} + v_{yy})
\end{align}
\end{subequations}
with $d_1,d_2=0.1$ and $\beta=1$. This set of equations generates a spiral wave formation, whose behavior can be approximately captured by two oscillating spatial modes. We apply our method to snapshots of $u(x,y,t)$ generated by the above equations. Snapshots are collected at discretized points of the $xy$-domain, resulting in a high-dimensional input data set with $n=10^4$.

We train the SINDy autoencoder with $d=2$. The resulting model is shown in Figure~\ref{fig:results}b. With two modes, the network discovers a model with nonlinear oscillatory dynamics. On test data, the fraction of unexplained variance of both the input data $\xv$ and the input derivatives $\dot{\xv}$ is $0.016$. The fraction of unexplained variance of $\dot{\zv}$ is $0.002$. Simulation of the dynamical model accurately captures the low dimensional dynamics, with the fraction of unexplained variance of $\zv$ totaling $1 \times 10^{-4}$.

\subsection{Nonlinear pendulum}

As a final example, we consider simulated video of a nonlinear pendulum in two spatial dimensions. The nonlinear pendulum is governed by the following second order differential equation:
\begin{equation}
  \ddot{z} = -\sin z.
\end{equation}
We simulate the system from several initial conditions and generate a series of snapshot images with a two-dimensional Gaussian centered at the center of mass, determined by the pendulum's angle $z$. This series of images is the high-dimensional data input to the autoencoder. Despite the fact that the position of the pendulum can be represented by a simple one-dimensional variable, methods such as PCA are unable to obtain a low-dimensional representation of this data set. A nonlinear autoencoder, however, is able to discover a one-dimensional representation of the data set.

For this example, we use a second-order SINDy model: that is, we use a library of functions including the first derivatives $\dot{\zv}$ to predict the second derivative $\ddot{\zv}$. This approach is the same as with a first order SINDy model but requires estimates of the second derivatives in addition to estimates of the first derivatives. Second order gradients of the encoder and decoder are therefore also required. Computation of the derivatives is discussed in Section~\ref{sec:computing_gradients}.

The SINDy autoencoder is trained with $d=1$. Of the ten training instances, five correctly identify the nonlinear pendulum equation. We calculate test error on trajectories from 50 randomly chosen initial conditions. The best model has a fraction of unexplained variance of $8 \times 10^{-4}$ for the decoder reconstruction of the input $\xv$. The fraction of unexplained variance of the SINDy model predictions for $\ddot{\xv}$ and $\ddot{\zv}$ are $3 \times 10^{-4}$ and $2 \times 10^{-2}$, respectively.

\section{Discussion}

We have presented a data-driven method for discovering interpretable, low-dimensional dynamical models and their associated coordinates for high-dimensional dynamical systems.  The simultaneous discovery of both is critical for generating dynamical models that are sparse, and hence interpretable and generalizable.
 Our approach takes advantage of the power of NNs by using a flexible autoencoder architecture to discover nonlinear coordinate transformations that enable the discovery of parsimonious, nonlinear governing equations. This work addresses a major limitation of prior approaches for the discovery of governing equations, which is that the proper choice of measurement coordinates is often unknown. We demonstrate this method on three example systems, showing that it is able to identify coordinates associated with parsimonious dynamical equations. For the examples studied, the identified models are interpretable and can be used for forecasting (extrapolation) applications (see Figure~\ref{fig:lorenz_si}).

A current limitation of our approach is the requirement for clean measurement data that is approximately noise-free. Fitting a continuous-time dynamical system with SINDy requires reasonable estimates of the derivatives, which may be difficult to obtain from noisy data. While this represents a challenge, approaches for estimating derivatives from noisy data such as the total variation regularized derivative can prove useful in providing derivative estimates \cite{chartrand_numerical_2011}.  Moreover, there are emerging NN architectures explicitly constructed for separating signals from noise~\cite{rudy2018deep}, which can be used as a pre-processing step in the data-driven discovery process advocated here.   Alternatively our method can be used to fit a discrete-time dynamical system, in which case derivative estimates are not required. Many methods for modeling dynamical systems work in discrete time rather than continuous time, making this a reasonable alternative. It is also possible to use the integral formulation of SINDy to abate noise sensitivity~\cite{Schaeffer2017pre}.

A major problem with deep learning approaches is that models are typically neither interpretable nor generalizable. Specifically, NNs trained solely for prediction may fail to generalize to classes of behaviors not seen in the training set. We have demonstrated an approach for using NNs to obtain classically interpretable models through the discovery of low-dimensional dynamical systems, which are well-studied and often have physical interpretations. Once the proper terms in the governing equations are identified, the discovered model can be generalized to study other parameter regimes of the dynamics. While the coordinate transformation learned by the autoencoder may not generalize to data regimes far from the original training set, if the dynamics are known, the network can be retrained on new data with fixed terms in the latent dynamics space. The problem of relearning a coordinate transformation for a system with known dynamics is greatly simplified from the original challenge of learning the correct form of the underlying dynamics without knowledge of the proper coordinate transformation.

The challenge of utilizing NNs to answer scientific questions requires careful consideration of their strengths and limitations. While advances in deep learning and computing power present a tremendous opportunity for new scientific breakthroughs, care must be taken to ensure that valid conclusions are drawn from the results. One promising strategy is to combine machine learning approaches with well-established domain knowledge: for instance physics-informed learning leverages physical assumptions into NN architectures and training methods. 
Methods that provide interpretable models have the potential to enable new discoveries in data-rich fields.  
This work introduced a flexible framework for using NNs to discover models that are interpretable from a standard dynamical systems perspective. In the future, this approach could be adapted using domain knowledge to discover new models in specific fields.

\section*{Acknowledgments}
This material is based upon work supported by the National Science Foundation Graduate Research Fellowship under Grant No. DGE-1256082. The authors also acknowledge support from the Defense Advanced Research Projects Agency (DARPA PA-18-01-FP-125) and the Army Research Office (ARO W911NF-17-1-0306 and W911NF-17-1-0422). This work was facilitated through the use of advanced computational, storage, and networking infrastructure provided by AWS cloud computing credits funded by the STF at the University of Washington. This research was funded in part by the Argonne Leadership Computing Facility, which is a DOE Office of Science User Facility supported under Contract DE-AC02-06CH11357.  We would also like to thank Jean-Christophe Loiseau and Karthik Duraisamy for valuable discussions about sparse dynamical systems and autoencoders.


\begin{spacing}{.9}
\small{
\setlength{\bibsep}{6.5pt}

}
\end{spacing}

\newpage

\beginsupplement

\section{Network Architecture and Training}

\subsection{Network architecture}\label{sec:si_architecture}
The autoencoder network consists of a series of fully-connected layers. Each layer has an associated weight matrix $\Wv$ and bias vector $\bv$. We use sigmoid activation functions $f(x) = 1/(1+\exp(-x))$, which are applied at all layers of the network, except for the last layer of the encoder and the last layer of the decoder. Other choices of activation function, such as rectified linear units and exponential linear units, may also be used and appear to achieve similar results.

\subsection{Loss function}\label{sec:si_loss_function}
The loss function used in training is a weighted sum of four terms: autoencoder reconstruction $\Lrecon$, SINDy prediction on the input variables $\Lsindyx$, SINDy prediction on the encoder variables $\Lsindyz$, and SINDy coefficient regularization $\Lreg$. For a data set with $m$ input samples, each loss is explicitly defined as follows:
\begin{subequations}
\begin{align}
  \Lrecon &= \frac{1}{m} \sum_{i=1}^m \left\| \xv_i - \psi(\varphi(\xv_i))\right\|_2^2 \\
  \Lsindyx  &= \frac{1}{m} \sum_{i=1}^m \left\|\dot{\xv}_i - \left(\nabla_\zv\psi(\varphi(\xv_i))\right)\left(\Thetav(\varphi(\xv_i)^T)\Xiv\right)\right\|_2^2 \\
  \Lsindyz &= \frac{1}{m} \sum_{i=1}^m \left\|\nabla_\xv\varphi(\xv_i)\dot{\xv}_i - \Thetav(\varphi(\xv_i)^T)\Xiv\right\|_2^2 \\
  \Lreg &= \frac{1}{pd} \left\| \Xiv \right\|_1.
\end{align}
\end{subequations}
The total loss function is
\begin{equation}
  \Lrecon + \lambda_1 \Lsindyx + \lambda_2 \Lsindyz + \lambda_3 \Lreg.
\end{equation}
$\Lrecon$ ensures that the autoencoder can accurately reconstruct the data from the intrinsic coordinates. $\Lsindyx$ and $\Lsindyz$ ensure that the discovered SINDy model captures the dynamics of the system by ensuring that the model can predict the derivatives from the data. $\Lreg$ promotes sparsity of the coefficients in the SINDy model.

\subsection{Computing derivatives}\label{sec:computing_gradients}
Computing the derivatives of the encoder variables requires propagating derivatives through the network. Our network makes use of an activation function $f(\cdot)$ that operates elementwise. Given an input $\xv$, we define the pre-activation values at the $j$th encoder layer as
\begin{equation}
   \lv_j = f(\lv_{j-1})\Wv_j + \bv_j.
\end{equation}
The first layer applies the weights and biases directly to the input so that
\begin{equation}
  \lv_0 = \xv \Wv_0 + \bv_0.
\end{equation}
The activation function is not applied to the last layer, so for an encoder with $L$ hidden layers the autoencoder variables are defined as
\begin{equation}
  \zv = f(\lv_{L-1})\Wv_L + \bv_L.
\end{equation}
Assuming that derivatives $d \xv/d t$ are available or can be computed, derivatives $d \zv/d t$ can also be computed:
\begin{align*}
  \frac{d \zv}{d t} &= \left(f'(\lv_{L-1})\circ \frac{d \lv_{L-1}}{d t}\right)\Wv_L
\end{align*}
with
\begin{align*}
  \frac{d \lv_j}{d t} &= \left( f'(\lv_{j-1}) \circ \frac{d \lv_{j-1}}{d t}\right)\Wv_j \\
  \frac{d \lv_0}{d t} &= \frac{d \xv}{d t}\Wv_0.
\end{align*}
For the nonlinear pendulum example, we use a second order SINDy model that requires the calculation of second derivatives. Second derivatives can be computed using the following:
\begin{align*}
  \frac{d^2 \zv}{dt^2} &= \left(  f''(\lv_{L-1}) \circ \frac{d \lv_{L-1}}{dt} \circ \frac{d \lv_{L-1}}{dt} + f'(\lv_{L-1}) \circ \frac{d^2 \lv_{L-1}}{dt^2} \right)\Wv_L \\
  \frac{d^2 \lv_j}{dt^2} &= \left(  f''(\lv_{j-1})\circ \frac{d \lv_{j-1}}{dt} \circ \frac{d\lv_{j-1}}{dt} + f'(\lv_{j-1}) \circ \frac{d^2 \lv_{j-1}}{dt^2} \right)\Wv_j \\
  \frac{d\lv_0}{dt} &= \frac{d^2\xv}{dt^2}\Wv_0.
\end{align*}

\subsection{Training procedure}\label{sec:si_training}
We train multiple models for each of the example systems. Each instance of training has a different random initialization of the network weights. The weight matrices $\Wv_j$ are initialized using the Xavier initialization: the entries are chosen from a random uniform distribution over $[-\sqrt{6/\alpha},\sqrt{6/\alpha}]$ where $\alpha$ is the dimension of the input plus the dimension of the output \cite{glorot2010understanding}. The bias vectors $\bv_j$ are initialized to 0 and the SINDy model coefficients $\Xiv$ are initialized so that every entry is 1. We train each model using the Adam optimizer for a fixed number of epochs \cite{DBLP:journals/corr/KingmaB14}. The learning rate and number of training epochs for each example are specified in Section~\ref{sec:si_results}.

To obtain parsimonious dynamical models, we use a sequential thresholding procedure that promotes sparsity on the coefficients in $\Xiv$, which represent the dynamics on the latent variables ${\bf z}$. Every 500 epochs, we set all coefficients in $\Xiv$ with a magnitude of less than $0.1$ to 0, effectively removing these terms from the SINDy model. This is achieved by using a mask $\Upsilonv$, consisting of 1s and 0s, that determines which terms remain in the SINDy model. Thus the true SINDy terms in the loss function are given by
\begin{equation}
  \lambda_1 \frac{1}{m} \sum_{i=1}^m \left\|\dot{\xv}_i - \left(\nabla_\zv\psi(\varphi(\xv_i))\right)\left(\Thetav(\varphi(\xv_i)^T)(\Upsilonv \circ \Xiv)\right)\right\|_2^2 + \lambda_2 \frac{1}{m} \sum_{i=1}^m \left\|\nabla_\xv\varphi(\xv_i)\dot{\xv}_i - \Thetav(\varphi(\xv_i)^T)(\Upsilonv \circ \Xiv)\right\|_2^2
\end{equation}
where $\Upsilonv$ is passed in separately and not updated by the optimization algorithm. Once a term has been thresholded out during training, it is permanently removed from the SINDy model. Therefore the number of active terms in the SINDy model can only be decreased as training continues. The $L_1$ regularization on $\Xiv$ encourages the model coefficients to decrease in magnitude, which combined with the sequential thresholding produces a parsimonious dynamical model.

While the $L_1$ regularization penalty on $\Xiv$ promotes sparsity in the resulting SINDy model, it also encourages nonzero terms to have smaller magnitudes. This results in a trade-off between accurately reconstructing the dynamics of the system and reducing the magnitude of the SINDy coefficients, where the trade-off is determined by the relative magnitudes of the loss weight penalties $\lambda_1,\lambda_2$ and the regularization penalty $\lambda_3$. The specified training procedure therefore typically results in models with coefficients that are slightly smaller in magnitude than those which would best reproduce the dynamics. To account for this, we add an additional coefficient refinement period to the training procedure. To perform this refinement, we lock in the sparsity pattern in the dynamics by fixing the coefficient mask $\Upsilonv$ and continue training for 1000 epochs without the $L_1$ regularization on $\Xiv$. This ensures that the best coefficients are found for the resulting SINDy model and also allows the training procedure to refine the encoder and decoder parameters. This procedure is analagous to running a debiased regression following the use of LASSO to select model terms \cite{tibshirani2015statistical}.

\subsection{Model selection}\label{sec:si_model_selection}
Random initialization of the NN weights is standard practice for deep learning approaches. This results in the discovery of different models for different instances of training, which necessitates comparison among multiple models. In this work, when considering the success of a resulting model, one must consider the parsimony of the SINDy model, how well the decoder reconstructs the input, and how well the SINDy model captures the dynamics.

To assess model performance, we calculate the fraction of unexplained variance of both the input data $\xv$ and its derivative $\dot{\xv}$. This error calculation takes into account both the decoder reconstruction and the fit of the dynamics. When considering parsimony, we consider the number of active terms in the resulting SINDy model. While parsimonious models are desirable for ease of analysis and interpretability, a model that is too parsimonious may be unable to fully capture the dynamics. In general, for the examples explored, we find that models with fewer active terms perform better on validation data (lower fraction of unexplained variance of $\dot{\xv}$) whereas models with more active terms tend to over-fit the training data. 

For each example system, we apply the training procedure to ten different initializations of the network and compare the resulting models. For the purpose of demonstration, for each example we show results for a chosen ``best'' model, which is taken to be the model with the lowest fraction of variance unexplained on validation data among models with the fewest active coefficients. While every instance of training does not result in the exact same SINDy sparsity pattern, the network tends to discover a few different closely related forms of the dynamics. We discuss the comparison among models for each particular example further in Section~\ref{sec:si_results}.

\section{Example Systems}\label{sec:si_results}

\subsection{Chaotic Lorenz system}\label{sec:si_lorenz}

\begin{figure*}
\centering
\includegraphics[width=\linewidth]{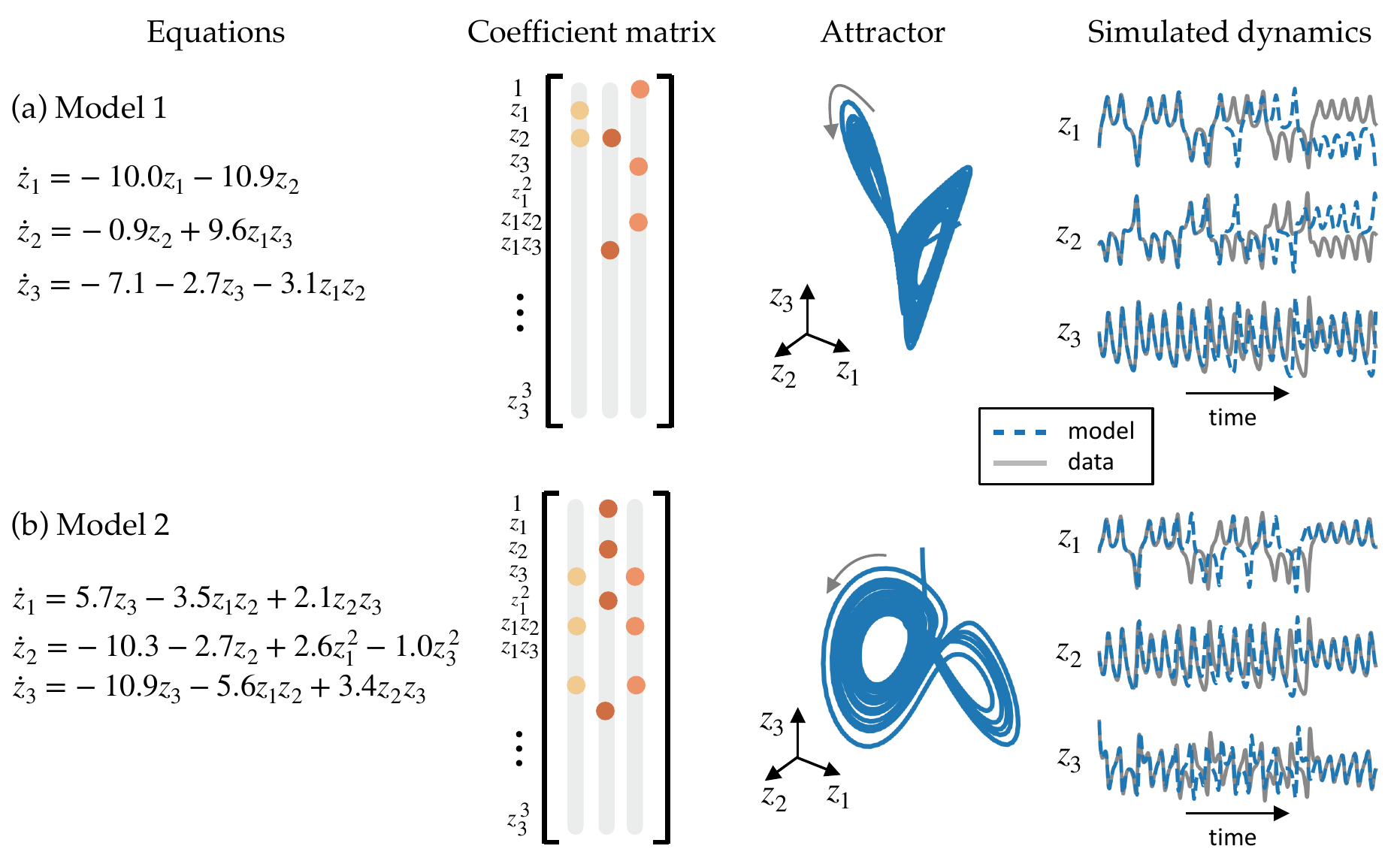}
\caption{Comparison of two discovered models for the Lorenz example system. For both models we show the equations, SINDy coefficients $\Xiv$, attractors, and simulated dynamics for two models discovered by the SINDy autoencoder. (a) A model with 7 active terms. This model can be rewritten in the same form as the original Lorenz system using the variable transformation described in Section~\ref{sec:si_lorenz}. Simulation of the model produces an attractor with a two lobe structure and is able to reproduce the true trajectories of the dynamics  for some time before eventually diverging due to the chaotic nature of the system. (b) A model with 10 active terms. The model has more terms than the true Lorenz system, but has a slightly lower fraction of unexplained variance of $\xv,\dot{\xv}$ than the model in (a). Simulation shows that the dynamics also lie on an attractor with two lobes. The model can accurately predict the true dynamics over a similar duration as (a).}
\label{fig:lorenz_si}
\end{figure*}

\begin{table}
\centering
\caption{Hyperparameter values for the Lorenz example}
\label{table:lorenz_params}
\begin{tabular}{|l|r|}\hline
\bf Parameter & \bf Value \\ \hline\hline
n & 128 \\
d & 3 \\
training samples & $5.12 \times 10^5$ \\
batch size & 8000 \\
activation function & sigmoid \\
encoder layer widths & $64,32$ \\
decoder layer widths & $32,64$ \\
learning rate & $10^{-3}$ \\
SINDy model order & 1 \\
SINDy library polynomial order & 3 \\
SINDy library includes sine & no \\
SINDy loss weight $\dot{\mathbf{x}}$, $\lambda_1$ & $10^{-4}$ \\
SINDy loss weight $\dot{\mathbf{z}}$, $\lambda_2$ & $0$ \\
SINDy regularization loss weight, $\lambda_3$ & $10^{-5}$\\ \hline
\end{tabular}
\end{table}

To create a high-dimensional data set with dynamics defined by the Lorenz system, we choose six spatial modes $\mathbf{u}_1,\dots,\mathbf{u}_6 \in \Rb^{128}$ and take
\begin{align*}
  \mathbf{x}(t) = \mathbf{u}_1 z_1(t) + \mathbf{u}_2 z_2(t) + \mathbf{u}_3 z_3(t) + \mathbf{u}_4 z_1(t)^3 + \mathbf{u}_5 z_2(t)^3 + \mathbf{u}_6 z_3(t)^3.
\end{align*}
where the dynamics of $\mathbf{z}$ are specified by the Lorenz equations
\begin{subequations}
\begin{align}
  \dot{z}_1 &= \sigma(z_2 - z_1) \\
  \dot{z}_2 &= z_1(\rho - z_3) - z_2 \\
  \dot{z}_3 &= z_1 z_2 - \beta z_3
\end{align}
\label{eq:si_lorenz}
\end{subequations}
with standard parameter values of $\sigma=10, \rho=28,\beta=8/3$. We choose our spatial modes $\mathbf{u}_1,\dots,\mathbf{u}_6$ to be the first six Legendre polynomials defined at 128 grid points on a 1D spatial domain $[-1,1]$. To generate our data set, we simulate the system from 2048 initial conditions for the training set, 20 for the validation set, and 100 for the test set. For each initial condition we integrate the system forward in time from $t=0$ to $t=5$ with a spacing of $\Delta t=0.02$ to obtain $250$ samples. Initial conditions are chosen randomly from a uniform distribution over $z_1 \in [-36,36]$, $z_2 \in [-48,48]$, $z_3 \in [-16,66]$. This results in a training set with 512,000 total samples.

Following the training procedure described in Section~\ref{sec:si_training}, we learn ten models using the single set of training data (variability among the models comes from the initialization of the network weights). The hyperparameters used for training are shown in Table~\ref{table:lorenz_params}. For each model we run the training procedure for $10^4$ epochs, followed by a refinement period of $10^3$ epochs. Of the ten models, two have 7 active terms, two have 10 active terms, one has 11 active terms, and five have 15 or more active terms. While all models have less than 1\% unexplained variance for both $\xv$ and $\dot{\xv}$, the three models with 20 or more active terms have the worst performance  predicting $\dot{\xv}$. The two models with 10 active terms have the lowest overall error, followed by models with 7, 15, and 18 active terms. While the models with 10 active terms have a lower overall error than the models with 7 terms, both have a very low error and thus we choose to highlight the model with the fewest active terms. A model with 10 active terms is shown in Figure~\ref{fig:lorenz_si} for comparison.

For analysis, we highlight the model with the lowest error among the models with the fewest active terms. The discovered model has equations
\begin{subequations}
\begin{align}
  \dot{z}_1 &= -10.0 z_1 - 10.9 z_2 \\
  \dot{z}_2 &= - 0.9 z_2 + 9.6 z_1 z_3 \\
  \dot{z}_3 &= -7.1 - 2.7 z_3 - 3.1 z_1 z_2.
\end{align}
\end{subequations}
While the structure of this model appears to be different from that of the original Lorenz system, we can define an affine transformation that gives it the same structure. The variable transformation $z_1 = \alpha_1 \tilde{z}_1$, $z_2 = \alpha_2 \tilde{z}_2$, $z_3 = \alpha_3 \tilde{z}_3 + \beta_3$ gives the following transformed system of equations:
\begin{subequations}
\begin{align}
  \dot{\tilde{z}}_1 &= \frac{1}{\alpha_1}\left(-10.0 \alpha_1 \tilde{z}_1 - 10.9 \alpha_2 \tilde{z}_2\right) \\
                    &= -10.0 \tilde{z}_1 - 10.9 \frac{\alpha_2}{\alpha_1}\tilde{z}_2 \\
  \dot{\tilde{z}}_2 &= \frac{1}{\alpha_2}\left(-0.9 \alpha_2 \tilde{z}_2 + 9.6 \alpha_1 \tilde{z}_1 (\alpha_3 \tilde{z}_3 + \beta_3)\right) \\
                    &= 9.6 \frac{\alpha_1}{\alpha_2}\beta_3\tilde{z}_1 - 0.9\tilde{z}_2 + 9.6 \frac{\alpha_1\alpha_3}{\alpha_2}\tilde{z}_1\tilde{z}_3 \\
  \dot{\tilde{z}}_3 &= \frac{1}{\alpha_3}\left(-7.1 - 2.7 (\alpha_3 \tilde{z}_3 + \beta_3) - 3.1 \alpha_1 \alpha_2 \tilde{z}_1\tilde{z}_2\right) \\
                    &= \frac{1}{\alpha_3}(-7.1 - 2.7 \beta_3) - 2.7 \tilde{z}_3 - 3.1 \frac{\alpha_1\alpha_2}{\alpha_3}\tilde{z}_1\tilde{z}_2.
\end{align}
\end{subequations}
By choosing $\alpha_1 = 1$, $\alpha_2 = -0.917$, $\alpha_3 = 0.524$, $\beta_3 = -2.665$, the system becomes
\begin{subequations}
\begin{align}
  \dot{\tilde{z}}_1 &= -10.0 \tilde{z}_1 + 10.0 \tilde{z}_2 \label{eq:lorenz_transformed1}\\
  \dot{\tilde{z}}_2 &= 27.7 \tilde{z}_1 - 0.9 \tilde{z}_2 - 5.5 \tilde{z}_1\tilde{z}_3 \label{eq:lorenz_transformed2}\\
  \dot{\tilde{z}}_3 &= -2.7 \tilde{z}_3 + 5.5 \tilde{z}_1\tilde{z}_2 \label{eq:lorenz_transformed3}.
\end{align}
\end{subequations}
This has the same form as the original Lorenz equations with parameters that are close in value, apart from an arbitrary scaling that affects the magnitude of the coefficients of $\tilde{z}_1\tilde{z}_3$ in \eqref{eq:lorenz_transformed2} and $\tilde{z}_1\tilde{z}_2$ in \eqref{eq:lorenz_transformed3}. The attractor dynamics for this system are very similar to the original Lorenz attractor and are shown in Figure~\ref{fig:lorenz_detail}c.

The learning procedure discovers a dynamical model by fitting coefficients that predict the continuous-time derivatives of the variables in a dynamical system. Thus it is possible for the training procedure to discover a model with unstable dynamics or which is unable to predict the true dynamics through simulation. We assess the validity of the discovered models by simulating the dynamics of the discovered low-dimensional dynamical system. Simulation of the system shows that the system is stable with trajectories existing on an attractor very similar to the original Lorenz attractor. Additionally, the discovered system is able to predict the dynamics of the original system. The fourth panel in Figure~\ref{fig:lorenz_si}a shows the trajectories found by stepping the discovered model forward in time as compared with the values of $\zv$ obtained by mapping samples of the high-dimensional data through the encoder. Although this is done on a new initial condition, the trajectories match very closely up to $t=5$, which is the duration of trajectories contained in the training set. After that the trajectories diverge, but the predicted trajectories remain on an attractor. The Lorenz dynamics are chaotic, and thus slight differences in coefficients or initial conditions cause trajectories to diverge quickly. 

For comparison, in Figure~\ref{fig:lorenz_si}b we show a second model discovered by the training procedure. This model has 10 active terms, as compared with 7 in the true Lorenz system. While the model contains additional terms not present in the original system, the dynamics lie on an attractor with a similar two lobe structure. Additionally, the system is able to predict the dynamics through simulation. This model has a lower error on test data than the original 7 term model, with a fraction of unexplained variance of $2\times 10^{-6}$ for $\xv$, $6 \times 10^{-5}$ for $\dot{\xv}$, and $3 \times 10^{-4}$ for $\dot{\zv}$.

\subsection{Reaction-diffusion}\label{sec:si_rd}

\begin{figure*}
\centering
\begin{overpic}[width=11.4cm]{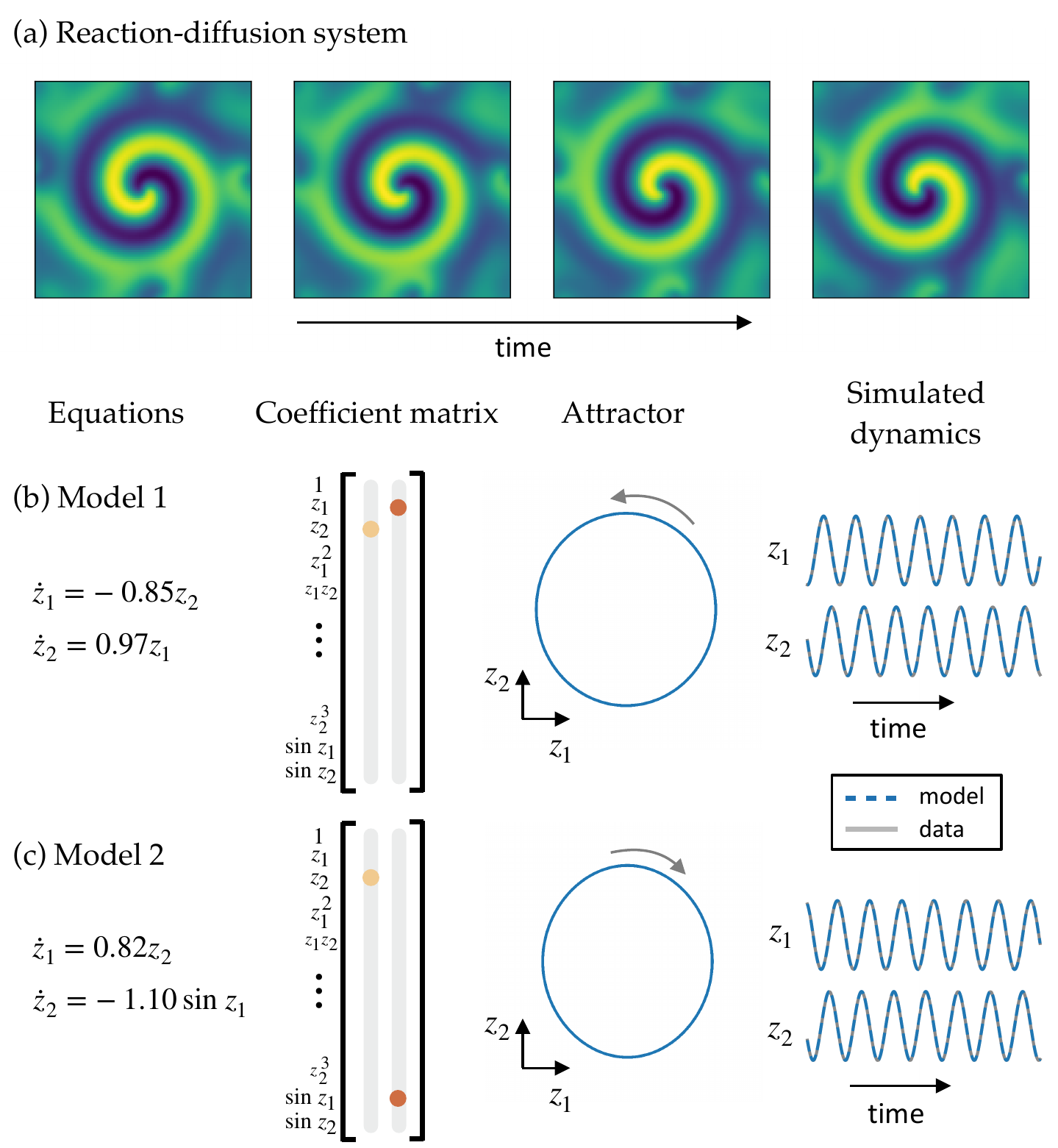}
\end{overpic}
\caption{Resulting models for the reaction-diffusion system. (a) Snapshots of the high-dimensional system show a spiral wave formation. (b,c) Equations, SINDy coefficients $\Xiv$, attractors, and simulated dynamics for two models discovered by the SINDy autoencoder. The model in (b) is a linear oscillation, whereas the model in (c) is a nonlinear oscillation. Both models achieve similar error levels and can predict the dynamics in the test set via simulation of the low-dimensional dynamical system.}
\label{fig:rd_detail}
\end{figure*}

\begin{table}
\centering
\caption{Hyperparameter values for the reaction-diffusion example}
\label{table:rd_params}
\begin{tabular}{|l|r|}\hline
\bf Parameter & \bf Value \\\hline\hline
n & $10^4$ \\
d & 2 \\
training samples & 8000 \\
batch size & 1024 \\
activation function & sigmoid \\
encoder layer widths & $256$ \\
decoder layer widths & $256$ \\
learning rate & $10^{-3}$ \\
SINDy model order & 1 \\
SINDy library polynomial order & 3 \\
SINDy library includes sine & yes \\
SINDy loss weight $\dot{\mathbf{x}}$, $\lambda_1$ & $0.5$ \\
SINDy loss weight $\dot{\mathbf{z}}$, $\lambda_2$ & $0.01$ \\
SINDy regularization loss weight, $\lambda_3$ & $0.1$\\ \hline
\end{tabular}
\end{table}

We generate data from a high-dimensional lambda-omega reaction-diffusion system governed by
\begin{subequations}
\begin{align}
  u_t &= (1-(u^2+v^2))u + \beta (u^2+v^2) v + d_1 (u_{xx} + u_{yy}) \\
  v_t &= -\beta (u^2+v^2)u + (1-(u^2+v^2))v + d_2 (v_{xx} + v_{yy})
\end{align}
\end{subequations}
with $d_1,d_2=0.1$ and $\beta=1$. The system is simulated from a single initial condition from $t=0$ to $t=10$ with a spacing of $\Delta t=0.05$ for a total of 10,000 samples. The initial condition is defined as
\begin{subequations}
\begin{align}
  u(y_1,y_2,0) &= \tanh\left(\sqrt{y_1^2 + y_2^2}\cos\left(\angle(y_1+iy_2) - \sqrt{y_1^2 + y_2^2}\right)\right) \\
  v(y_1,y_2,0) &= \tanh\left(\sqrt{y_1^2 + y_2^2}\sin\left(\angle(y_1+iy_2) - \sqrt{y_1^2 + y_2^2}\right)\right)
\end{align}
\end{subequations}
over a spatial domain of $y_1 \in [-10,10],\ y_2\in[-10,10]$ discretized on a grid with 100 points on each spatial axis. The solution of these equations results in a spiral wave formation. We apply our method to snapshots of $u(y_1,y_2,t)$ generated by the above equations, multiplied by a Gaussian $f(y_1,y_2) = \exp(-0.1(y_1^2+y_2^2))$ centered at the origin to localize the spiral wave in the center of the domain. Our input data is thus defined as $\xv(t) = f(:,:) \circ u(:,:,t) \in \Rb^{10^4}$. We also add Gaussian noise with a standard deviation of $10^{-6}$ to both $\xv$ and $\dot{\xv}$. Four time snapshots of the input data are shown in Figure~\ref{fig:rd_detail}a.

We divide the total number of samples into training, validation, and test sets: the last 1000 samples are taken as the test set, 1000 samples are chosen randomly from the first 9000 samples as a validation set, and the remaining 8000 samples are taken as the training set. We train ten models using the procedure outlined in Section~\ref{sec:si_training} for $3 \times 10^3$ epochs followed by a refinement period of $10^3$ epochs. Hyperparameters used for training are shown in Table~\ref{table:rd_params}. Nine of the ten resulting dynamical systems models have two active terms and one has three active terms. The dynamical equations, SINDy coefficient matrix, attractors, and simulated dynamics for two example models are shown in Figure~\ref{fig:rd_detail}b,c. The models with two active terms all have one of the two forms shown in the figure: three models have a linear oscillation and six models have a nonlinear oscillation. Both model forms have similar levels of error on the test set and are able to predict the dynamics in the test set from simulation, as shown in the fourth panel of Figure~\ref{fig:rd_detail}b,c.

\subsection{Nonlinear pendulum}\label{sec:si_pendulum}

\begin{figure*}
\centering
\begin{overpic}[width=\linewidth]{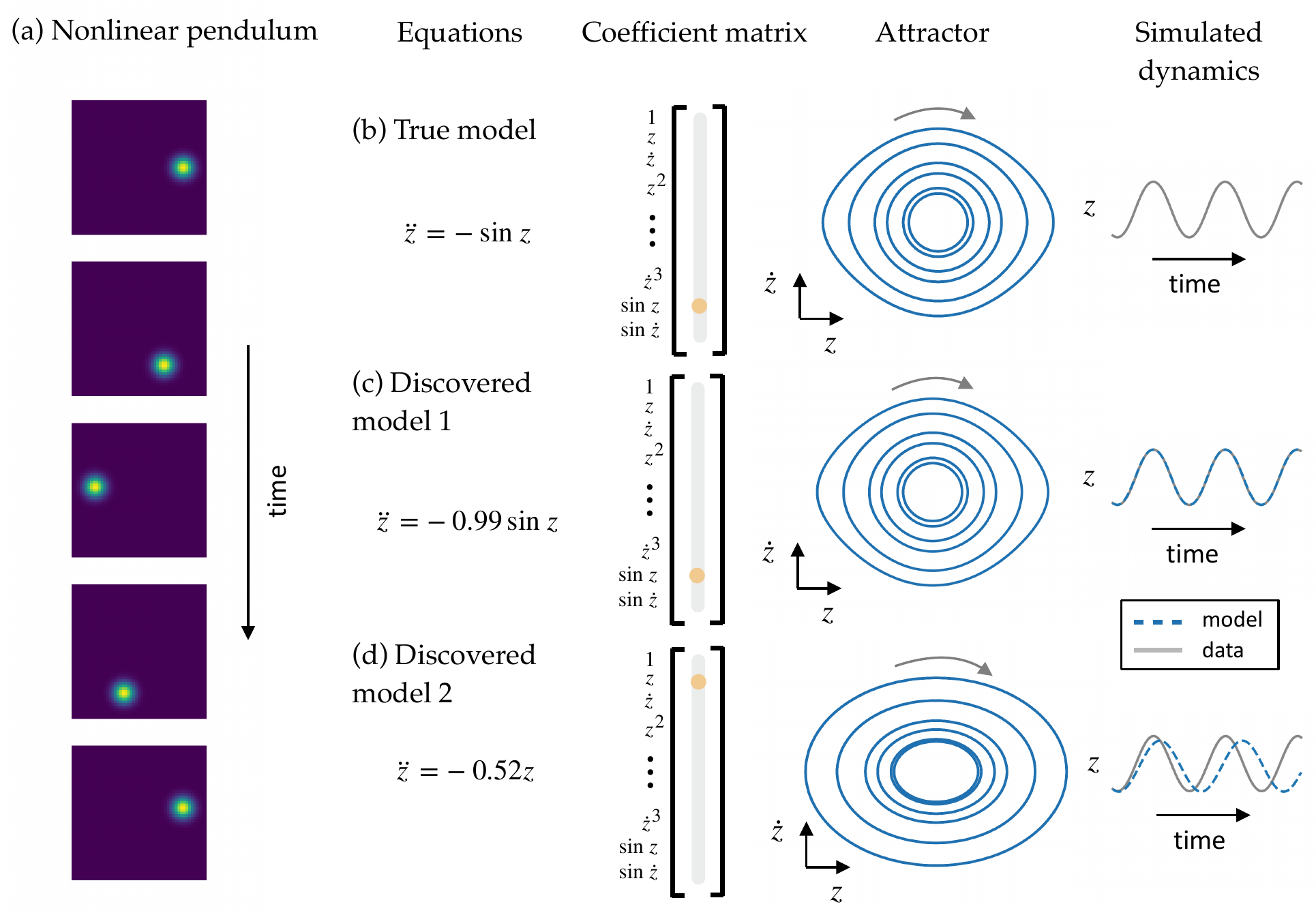}
\end{overpic}
\caption{Resulting models for the nonlinear pendulum. (a) Snapshots of the high-dimensional system are images representing the position of the pendulum in time. (b,c,d) Comparison of two discovered models with the true pendulum dynamics. Equations, SINDy coefficients $\Xiv$, attractors, and simulated dynamics for the true pendulum equation are shown in (b). The model in (c) correctly discovered the true form of the pendulum dynamics. Both the image of the attractor and simulations match the true dynamics. (d) In one instance of training, the SINDy autoencoder discovered a linear oscillation for the dynamics. This model achieves a worse error than the model in (c).}
\label{fig:pendulum_detail}
\end{figure*}

\begin{table}
\centering
\caption{Hyperparameter values for the nonlinear pendulum example}
\label{table:pendulum_params}
\begin{tabular}{|l|r|}\hline
\bf Parameter & \bf Value \\\hline\hline
n & $2601$ \\
d & 1 \\
training samples & $5 \times 10^4$\\
batch size & 1024 \\
activation function & sigmoid \\
encoder layer widths & $128,64,32$ \\
decoder layer widths & $32,64,128$ \\
learning rate & $10^{-4}$ \\
SINDy model order & 2 \\
SINDy library polynomial order & 3 \\
SINDy library includes sine & yes \\
SINDy loss weight $\dot{x}$, $\lambda_1$ & $5 \times 10^{-4}$ \\
SINDy loss weight $\dot{z}$, $\lambda_2$ & $5 \times 10^{-5}$ \\
SINDy regularization loss weight, $\lambda_3$ & $10^{-5}$\\\hline
\end{tabular}
\end{table}

The nonlinear pendulum equation is given by
\begin{equation}
  \ddot{z} = -\sin z. \label{eq:pendulum}
\end{equation}
We generate synthetic video of the pendulum in two spatial dimensions by creating high-dimensional snapshots given by
\begin{equation}
  x(y_1,y_2,t) = \exp\left(-\!20\!\left((y_1\! -\! \cos(z(t)\!-\!\pi/2))^2 + (y_2\! -\! \sin(z(t)\!-\!\pi/2))^2 \right) \right)
\end{equation}
at a discretization of $y_1,y_2\!\in\![-1.5,1.5]$. We use 51 grid points in each dimension resulting in snapshots $\xv(t) \in \Rb^{2601}$. To generate a training set, we simulate \eqref{eq:pendulum} from 100 randomly chosen initial conditions with $z(0) \in [-\pi,\pi]$ and $\dot{z}(0) \in [-2.1,2.1]$. The initial conditions are selected from a uniform distribution in the specified range but are restricted to conditions for which the pendulum does not have enough energy to do a full loop. This condition is determined by checking that $|\dot{z}(0)^2/2 - \cos z(0)| \leq 0.99$.

Following the training procedure outlined in Section~\ref{sec:si_training}, we train ten models for $5 \times 10^3$ epochs followed by a refinement period of $10^3$ epochs. Hyperparameters used for this example are shown in Table~\ref{table:pendulum_params}. Five of the ten resulting models correctly recover the nonlinear pendulum equation. These five models have the best performance of the ten models. The attractor and simulated dynamics for the best of these five models are shown in Figure~\ref{fig:pendulum_detail}. One model, also shown in Figure~\ref{fig:pendulum_detail}, recovers a linear oscillator. This model is able to achieve a reasonably low prediction error for $\ddot{\xv},\ddot{\zv}$ but the simulated dynamics, while still oscillatory, appear qualitatively different from the true pendulum dynamics. The four remaining models all have two active terms in the dynamics and have a worse performance than the models with one active term.

\end{document}